\documentclass[aps,prl,superscriptaddress,reprint]{revtex4-1}
\usepackage{amsmath}
\usepackage{amssymb}
\usepackage{graphicx}
\usepackage[colorlinks,urlcolor=blue]{hyperref}
\usepackage{url}
\usepackage{color,xcolor}
\usepackage{ulem}
\usepackage{float}
\usepackage{epstopdf}
\usepackage{appendix}
\begin{document}

\title{Similarities and differences between nickelate and cuprate films \\
grown on a SrTiO$_3$ substrate}
\author{Yang Zhang}
\author{Ling-Fang Lin}
\affiliation{Department of Physics and Astronomy, University of Tennessee, Knoxville, Tennessee 37996, USA}
\affiliation{School of Physics, Southeast University, Nanjing 211189, China}
\author{Wenjun Hu}
\affiliation{Department of Physics and Astronomy, University of Tennessee, Knoxville, Tennessee 37996, USA}
\author{Adriana Moreo}
\affiliation{Department of Physics and Astronomy, University of Tennessee, Knoxville, Tennessee 37996, USA}
\affiliation{Materials Science and Technology Division, Oak Ridge National Laboratory, Oak Ridge, Tennessee 37831, USA}
\author{Shuai Dong}
\affiliation{School of Physics, Southeast University, Nanjing 211189, China}
\author{Elbio Dagotto}
\affiliation{Department of Physics and Astronomy, University of Tennessee, Knoxville, Tennessee 37996, USA}
\affiliation{Materials Science and Technology Division, Oak Ridge National Laboratory, Oak Ridge, Tennessee 37831, USA}

\date{\today}

\begin{abstract}
The recent discovery of superconductivity in Sr-doped NdNiO$_2$ films grown on SrTiO$_3$ started a novel field within unconventional superconductivity. To understand the similarities and differences between nickelate and cuprate layers on the same SrTiO$_3$ substrate, here based on the density functional theory we have systematically investigated the structural, electronic, and magnetic properties of NdNiO$_2$/SrTiO$_3$ and CaCuO$_2$/SrTiO$_3$ systems. Our results revealed a strong lattice reconstruction in the case of NdNiO$_2$/SrTiO$_3$, resulting in a polar film, with the surface and interfacial NiO$_2$ layers presenting opposite displacements. To avoid the ``polar catastrophe'', the NiO$_2$ surface to the vacuum reconstructs as well.  However, for CaCuO$_2$/SrTiO$_3$, the distortions of those same two CuO$_2$ layers were in the same direction. In addition, we found this distortion to be approximately independent of the studied range of film thickness for the nickelate films. Furthermore, we also observed a two-dimensional electron gas at the interface between NdNiO$_2$ and SrTiO$_3$, caused by the polar discontinuity, in agreement with recent literature. For NdNiO$_2$/SrTiO$_3$ the two-dimensional electron gas extends over several layers, while for CaCuO$_2$/SrTiO$_3$ this electronic rearrangement is very localized at the interface between CaCuO$_2$ and SrTiO$_3$. The electronic reconstruction found at the interface involves a strong occupation of the Ti $3d_{xy}$ state. In both cases, there is a significant electronic charge transfer from the surface Ni or Cu layers to the Ti interface layer. The interfacial Ni and Cu layer is hole and electron doped, respectively. By introducing magnetism and electronic correlation, we observed that the $d_{3z^2-r^2}$ orbital of Ni becomes itinerant while the same orbital for Cu remains doubly occupied, establishing a clear two- vs one-orbital active framework for the description of these systems. Furthermore, we also observed a strong magnetic reconstruction at the NdNiO$_2$ surface to vacuum layer where magnetism is basically suppressed.

\end{abstract}

\maketitle
\section{I. Introduction}
Since the discovery of superconductivity in the cuprates~\cite{Bednorz:Cu}, two-dimensional (2D) correlated systems with transition metal square lattice structures became the focus of research to study unconventional superconductors and potentially to continue increasing the critical temperature~\cite{Dagotto:rmp94,Stewart:Rmp,Scalapino:Rmp,Dagotto:Rmp}. For this reason, the recent discovery of  superconductivity in the infinite-layer Sr-doped NdNiO$_2$ (NNO) attracted
much attention~\cite{Li:Nature,Nomura:Prb,Botana:prx,Li:prl20,Hepting:nm}. The discovery was reported employing
films of NNO grown on SrTiO$_3$ (STO) \cite{Li:Nature}, with a critical temperature $T_c$ $\sim 15$~K.
NNO displays a typical quasi-two-dimensional square lattice structure with the space group $P4/mmm$ (No. $123$) \cite{Li:Nature}, similar to the Cu-based CaCuO$_2$ (CCO) \cite{Azuma:Nature}. Since the $3d$ electronic occupied configuration of Ni$^{\rm 1+}$ ($d^9$) is isoelectronic with Cu$^{\rm 2+}$ ($d^9$), the superconducting mechanism of infinite-layer nickelate was expected to be similar to CCO, where superconductivity was found upon hole doping \cite{Azuma:Nature}. Many theoretical efforts revealed similarities and differences between individual infinite-layer nickelate and cuprates in bulk form~\cite{Botana:prx,Sakakibara:arxiv,Jiang:prl,Wu:prb,Zhang:prb,Werner:prb,Gu:prb,Karp:prx,Kitatani:arxiv}.

However, recent developments indicate that superconductivity is absent in bulk Sr-doped NNO~\cite{Li:cm,wang:arxiv}. These results suggest that the existence of an interface NNO-STO may be crucial to understand the superconductivity observed in the infinite-layer nickelate film. Moreover, superconductivity was also reported at the interface between CCO and STO~\cite{Aruta:prb,Castro:prl}. This is all reminiscent of previous results for the LaAlO$_3$(LAO)/STO interface where superconductivity was observed experimentally and attributed to an interfacial two-dimensional electron gas (2DEG) driven by the polar discontinuity~\cite{Thiel:sci,Reyren:sci,Dagotto:nature}. Related STO perovskite interfaces have also been widely studied theoretically, leading to interesting predictions at those interfaces~\cite{Piskunov:ssi,Piskunov:pssc}. In addition, there are many efforts to engineer charge transfer and orbital polarization in LaTiO$_3$/LaNiO$_3$ superlattices to obtain superconductivity \cite{Chen:prl13-1,Chen:prl13-2,Disa:prl}. In this superlattice structure, due to carriers doping, a single $3d_{x^2-y^2}$ orbital was obtained at band edges, providing a general method for designing new superconductors.

It is surprising that a conducting system would emerge at the interface of two wide-gap insulating oxides \cite{Ohtomo:nature,Ohtomo:nature04}. In the LAO/STO system, it was found that the 2DEG would become magnetic and superconducting at low temperatures~\cite{Brinkman:nm,Reyren:sci}. One possible explanation relies on the polar catastrophe that forces an interfacial reconstruction~\cite{Ohtomo:nature04,Nakagawa:nm}.  Furthermore, in the LAO/STO interface, the Ti $3d_{xy}$ orbital plays an important role for the interface ferromagnetism (FM) and superconductivity, produced by such electronic reconstruction~\cite{Dikin:Prl,Pavlenko:Prb}. These interesting developments in LAO/STO, and the absence of superconductivity in bulk NNO, provides the framework for the present work, as well as recent efforts by other groups.
Can a similar interesting interfacial phenomenon be crucial for the physics of NNO/STO? In particular, do nickelates and cuprates films display similar or different behavior when grown on a STO substrate?

In the infinite-layer nickelate, the system alternatively stacks the highly charged (Nd)$^{\rm 3+}$ and (NiO$_2$)$^{\rm 3-}$ layers along the [001] direction, while the STO substrate contains charge-neutral layers (SrO)$^0$ and (TiO$_2$)$^0$. This configuration would produce a net electric field  inside the nickelate component, resulting from an electric potential that diverges as the NNO thickness grows. To avoid this polar catastrophe, an electronic reconstruction occurs and hole doping [from Ni$^{\rm 1+}$ ($d^9$) to Ni$^{\rm 2+}$ ($d^8$)] at the interface could be expected, leading to an oscillating electric potential that remains finite. The possibility of ``polar catastrophe'' physics in the NNO/STO heterojunction was recently theoretically discussed in Ref.~\cite{He:prb}.

To better understand the similarities and differences between nickelates' and cuprates' films grown on an STO substrate, here using density functional theory (DFT) we provide a comprehensive first-principles study of NNO/STO and CCO/STO systems. Our main results are: (1) We found a strong lattice reconstruction in NNO/STO resulting in a polar nickelate film, with the limiting surface and interfacial NiO$_2$ layers presenting opposite displacements. On the other hand, in CCO/STO the distortions of those same two CuO$_2$ layers were in the same direction. Within our accuracy, the atomic reconstruction was basically independent of the thickness of the NNO or CCO films. (2) We also observed a strong electronic reconstruction for the two systems. Remarkably, and in agreement
with our qualitative expectation, a 2DEG develops at the interfacial
layer between NNO and STO due to the polar discontinuity. This electron gas extends over several layers beyond the interface. In CCO/STO, however, the electronic rearrangement was far more constrained in size and only observed at the interface between CCO and STO. The hole-doping at the interfacial NiO$_2$ layer was observed, but the interfacial CuO$_2$ layer was found to be electron doped. By comparing the unrelaxed and relaxed slab models, we believe that the 2DEG was indeed induced by the polar discontinuity. Furthermore, a strong occupation of the Ti $3d_{xy}$ orbital was observed at the TiO$_2$ interface for both NNO/STO and CCO/STO. At the interfacial TiO$_2$ layer, the Ti $3d_{xy}$ orbital becomes occupied due to its energy being lowered by $\sim$ $-0.58$ eV at the $\Gamma$ point. In the LaAlO$_3$/SrTiO$_3$ interface, the Ti $3d_{xy}$ orbital is important for the observed interfacial FM and superconductivity~\cite{Dikin:Prl,Pavlenko:Prb}. Thus, it is reasonable to assume the strongly occupied Ti $3d_{xy}$ orbital may be associated to the observed superconductivity of the infinite-layer nickelates as well. By introducing magnetism in the calculations, the $d_{3z^2-r^2}$ orbital of Ni becomes itinerant, probably contributing to the non-magnetic properties of NNO/STO, while the $d_{3z^2-r^2}$ orbital of Cu remains doubly occupied and thus not relevant at the dopings considered experimentally, establishing another qualitative difference between the two cases. In addition, the magnetism of the NdNiO$_2$ surface layer was basically suppressed.

\begin{figure}
\centering
\includegraphics[width=0.48\textwidth]{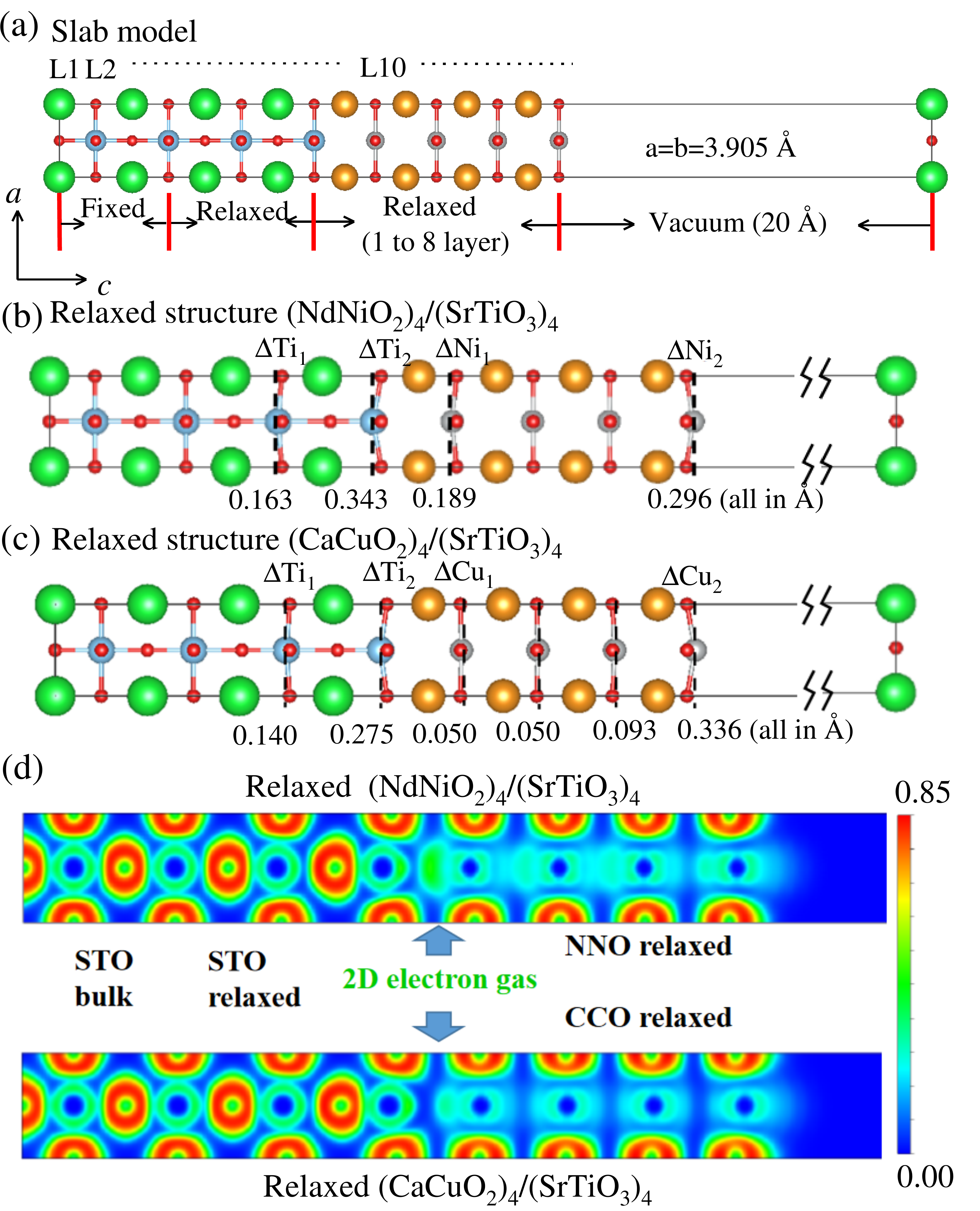}
\caption{(a) Schematic unrelaxed slab model of (NNO)$_n$/(STO)$_4$ with the convention: Orange = Nd or Ca; Gray = Ni or Cu; Red = O; Green = Sr; Blue = Ti. Here, we fixed $a=b=3.905$~\AA. We considered two undistorted STO layers (namely with atomic positions fixed) to simulate the effects of the rest of the STO substrate. (b,c) Schematic relaxed slab model of (NNO)$_4$/(STO)$_4$  and (CCO)$_4$/(STO)$_4$, respectively. (d) Electron localization function of the relaxed superlattice for (NNO)$_4$/(STO)$_4$  and (CCO)$_4$/(STO)$_4$.}
\label{Fig1}
\end{figure}

\section{II. Method and Model system}

In contrast to the insulator cuprates, NNO displays metallic characteristics \cite{Li:Nature}. Studying the superconducting dome of (Nd,Sr)NiO$_2$, only a weakly insulating behavior was reported on both sides of the dome \cite{Li:prl20}. STO is a wide-band semiconductor. To simulate the infinite-layer nickelates and cuprates grown on STO [001], we constructed the slab models using (NNO)$_n$/(STO)$_4$ and (CCO)$_n$/(STO)$_4$ superlattices ($n = 1-8$ are the layers of nickelates or cuprates film specifically analyzed here, corresponding to $1-8$ unit cells). To match the STO [001] substrate, the in-plane lattice constants were fixed as $a=b=3.905$~\AA, the STO values.  Here, a $1\times1$ lattice in the $xy$ plane was adopted without considering the octahedral rotations. These rotations could affect the STO bandwidth and should have an influence on the precise amount of charge transfer. However, those rotations will not change the tendency towards charge transfer since this is caused by the polar catastrophe. Furthermore, the octahedral rotations almost disappear at the TiO$_2$ interfacial layer~\cite{Geisler:prb}. Hence, to save computing resources, we have not introduced octahedral rotations since we focused on the qualitative phenomenon of charge transfer rather than on precise values of what amount of charge is transferred.

In the present study, we perform first-principles DFT calculations using the projector augmented wave (PAW) pseudopotentials with the Perdew-Burke-Ernzerhof (PBE) format, as implemented in the Vienna {\it ab initio} Simulation Package (VASP) code~\cite{Kresse:Prb,Kresse:Prb96,Blochl:Prb,Perdew:Prl}. We fixed the atomic positions of two STO layers to their bulk values [see Fig.~\ref{Fig1}(a)] while all other inner atomic positions were fully relaxed until the Hellman-Feynman force on each atom was smaller than 0.01 eV/~\AA.  The plane-wave cutoff energy was $550$~eV and the Nd $4f$ electrons are considered frozen in the core~\cite{Nomura:Prb}. The mesh was appropriately modified for both the non-magnetic and magnetic calculations to render the $k$-point densities approximately the same in reciprocal space (i.e. $15\times15\times1$ for the NM configuration).

Before studying heterostructures, we checked the physical properties of individual NNO, CCO, and STO, respectively~\cite{bandgap}. Based on our structural optimization calculation with PBE potential, the optimized crystal lattices of NNO are $a=b=3.907$~\AA ~and $c=3.304$~\AA, which are closed to the experimental bulk values ($a=b=3.921$~\AA ~and $c=3.281$~\AA) \cite{Li:Nature}. We also obtained the lateral lattice constants of STO ($a=b=3.943$~\AA) that are also in agreement with experimental values ($a=b=3.905$~\AA)~\cite{Mitsui:sto}. Moreover, the optimized crystal lattices of CCO were found to be $a=b=3.874$~\AA ~and $c=3.205$~\AA, in good agreement with the corresponding bulk values ($a=b=3.855$~\AA ~and $c=3.181$~\AA)~\cite{Azuma:Nature}.

Then, we also calculated their electronic structures and found them to be consistent with previous calculations~\cite{Botana:prx,Nomura:Prb,Wu:prb,Karp:prx}. The detailed results of bulk electronic structures can be found in Appendix A. Furthermore, STO is known to be located at the vicinity between paraelectric and ferroelectric phases, rendering it sensitive to artificial strain. By fixing the crystal lattice of STO ($a=b=3.905$~\AA), then the relaxation procedure produced a value $c=4.032$~\AA ~that may lead to an artificial distortion because STO wishes to expand its lattice along the $c$-axis. To verify this point, we also fixed the in-plane lattice to the optimized STO constants ($a=b=3.943$~\AA) and we found this procedure does not change the main conclusions (see Appendix B).

\section{III. Results}

\subsection{A. Atomic reconstruction}
As illustrated in Fig.~\ref{Fig1}(b) for $n = 4$, both the surface and interfacial NiO$_2$ layers bend after structural relaxation, and the corresponding Ni ions move towards {\it different} directions with displacements
$\Delta= 0.296$ and $-0.189$~\AA. However, because the displacements are different in magnitude the net result is a polar film. The displacements of the Ni ions for the two central NiO$_2$ layers are much smaller. Furthermore, the interfacial Ti ions of the STO substrate show ferroelectric-like displacements away from the NNO/STO interface with values $\Delta= -0.343$ and $-0.163$~\AA, respectively. Note that STO is known to be located at the vicinity between paraelectric and ferroelectric phases. Thus, the polar displacements of Ti may affect by the ferroelectric distortion of STO itself (this issue deserves further investigation beyond the scope of this publication). For the CCO/STO system, the Ti atoms also display similar ferroelectric-like distortions with similar displacements $\Delta= -0.275$ and $-0.140$~\AA. However, the optimized CCO/STO superlattice indicates that the distortions of each CuO$_2$ layer are in the {\it same} direction along the $c$-axis, as shown in Fig.~\ref{Fig1}(c). This results in a polar film as well. Polar distortions similar to those of CCO/STO observed here were also reported for LAO layers grown on STO [001] ~\cite{Pentcheva:Prl}.

\begin{figure}
\centering
\includegraphics[width=0.48\textwidth]{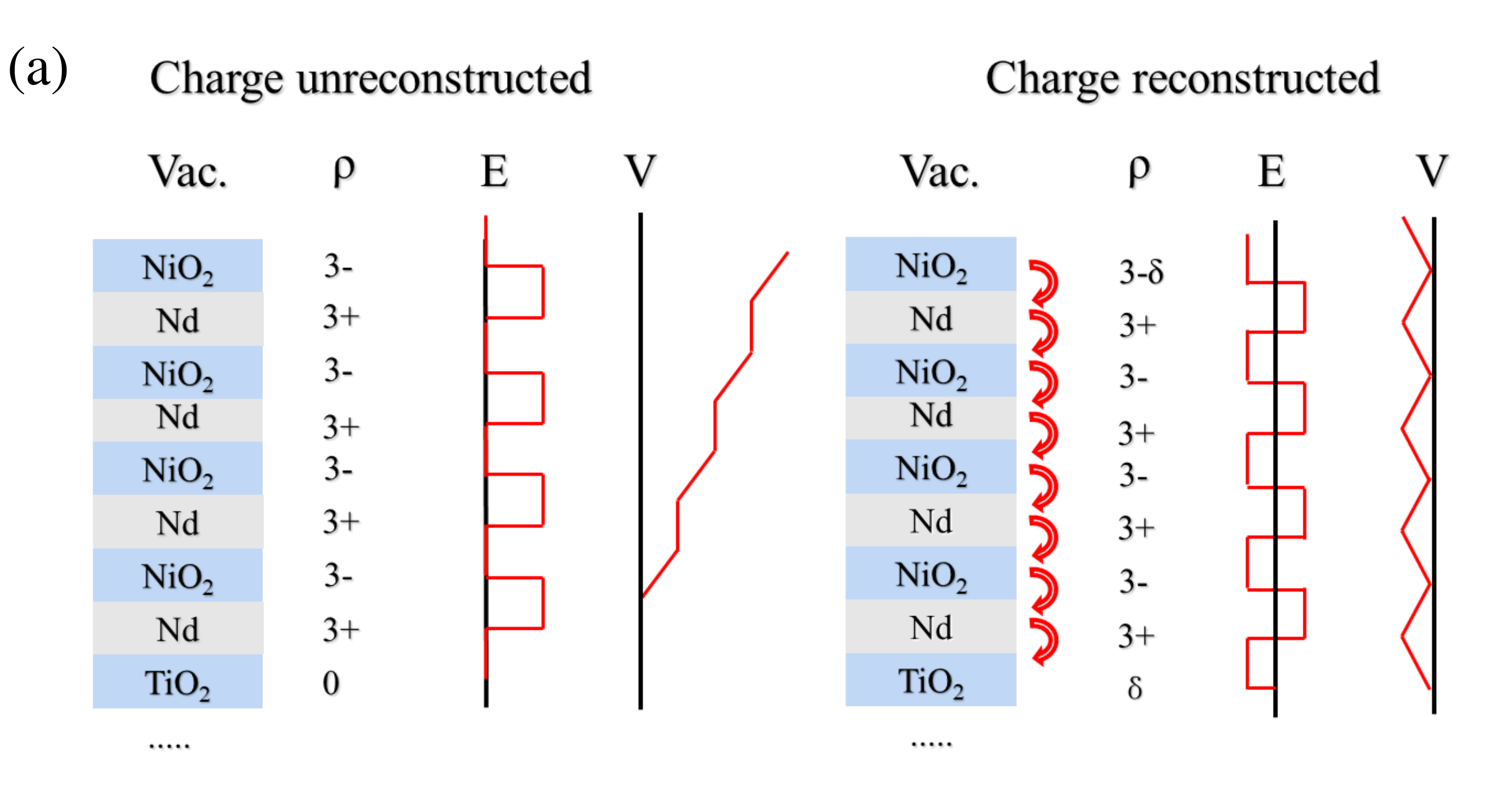}
\includegraphics[width=0.48\textwidth]{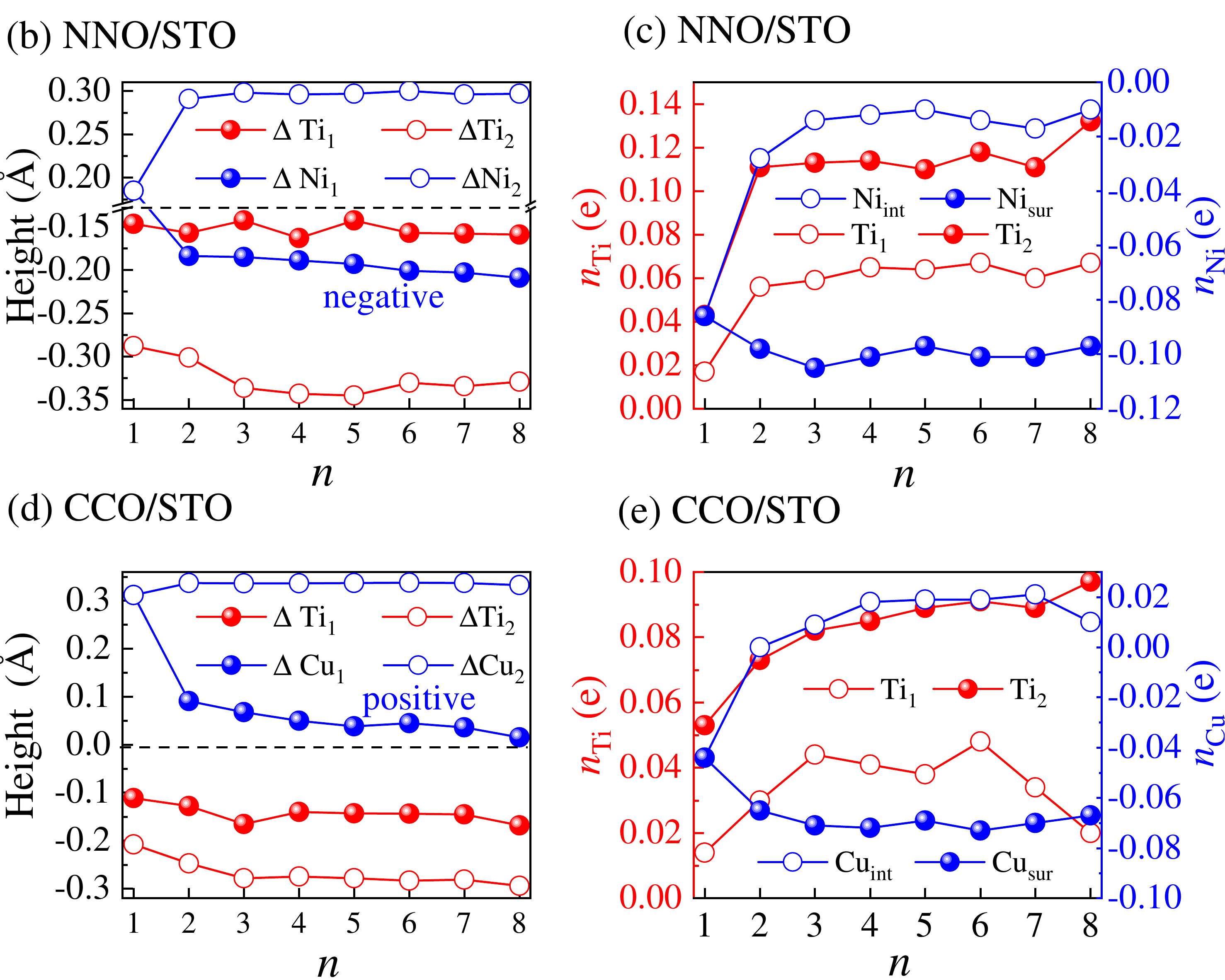}
\caption{(a) Possible polar catastrophe illustration predicted for (NdNiO$_2$)$_n$/SrTiO$_3$. (b) The displacements along the $c$-axis for the four layers of Ni or Ti atoms corresponding to different superlattices (NNO)$_n$/STO ($n = 1-8$). In Fig.~\ref{Fig1}~(a) the notation $\Delta M$ ($M$=Ni$_1$, Ni$_2$, Cu$_1$, Cu$_2$, Ti$_1$ and Ti$_2$) was introduced. (c) The charge transfer of Ni and Ti atoms for different (NNO)$_n$/(STO)$_4$ ($n = 1-8$), compared with the corresponding bulk values. (d) The main displacements along the $c$-axis for Cu or Ti atoms for different (CCO)$_n$/(STO)$_4$ ($n = 1-8$). (e) The charge transfer of Ni and Ti sites for different (CCO)$_n$/(STO)$_4$ ($n = 1-8$), compared with corresponding bulk values. In (c,d) the subindex "int" ("sur") denotes interface (surface). Note that the single $M$O$_2$ ($M$=Ni or Cu) layer is both interface and surface for $n = 1$ case.}
\label{Fig2}
\end{figure}

Those atomic displacements are relevant within the so-called ``polar catastrophe''. When growing the highly charged (Nd)$^{\rm 3+}$ and (NiO$_2$)$^{\rm 3-}$ layers on the charge-neutral STO substrate, it will induce an ever-increasing dipole strength as the number of layers grows if the ionic charges are maintained at its normal value. In this case, a huge linear electrostatic potential across the NNO layers will cause the so-called polar catastrophe, as shown in Fig.~\ref{Fig2}(a). To avoid this phenomenon, the surfaces are expected to be reconstructed (see Fig.~\ref{Fig2}(a)). In particular, an electronic reconstruction from Ni$^{\rm 1+}$ ($d^9$) to Ni$^{\rm 2+}$ ($d^8$) could be expected at the surface. As a result, the surface-vacuum polar distortion is induced by the polar field caused by the charge transfer. Namely, the lattice distortion is driven by the polar nature of the surface.

To better understand the effect of the film thickness for NNO and CCO, we fully relaxed the slab model using different number of NNO and CCO layers ($n = 1-8$, corresponding to $1-8$ unit cells). We observed that the distortion of the atomic layer reaches a value that is approximately independent of the film thickness for NNO/STO, at least within the range we studied. For CCO/STO, the distortion at the interfacial CuO$_2$ gradually reduces its value with increasing the film thickness. Moreover, as shown in Figs.~\ref{Fig2}(b) and (d), we observed that the thickness of the NNO and CCO layers does not alter the main qualitative results: (1) the two relaxed TiO$_2$ layers present a uniform polar distortion direction along the $c$-axis both for NNO and CCO; (2) the displacements of the surface and interfacial NiO$_2$ layers are opposite to each other, while the distortions of the surface and interfacial CuO$_2$ layers are in the same direction. In addition, the values of the displacements do not change substantially by increasing the film thickness of the NNO and CCO layers. We summarized the main displacements along the $c$-axis in Figs.~\ref{Fig2}(b) and (d). Note the prominent
difference NNO vs CCO with regards to the negative $\Delta$Ni$_1$ vs the positive $\Delta$Cu$_1$.

Furthermore, we also calculated the $3d$ electronic density for both (NNO)$_n$/(STO)$_4$ and (CCO)$_n$/(STO)$_4$ superlattices and compared with the corresponding bulk systems, see Figs.~\ref{Fig2}(b) and (d). For (NNO)$_n$/(STO)$_4$, by comparing with the bulk, it is clear that the relaxed two Ti sites gathered additional electrons while the surface Ni sites lost electrons, indicating the transfer of charge from the Ni surface to the interface with STO. Based on the DFT results \cite{chargetransfer}, we estimate that for the case $n = 4$ the Ni surface lost charge by an amount $-0.101$~e/Ni, while the interface Ti gained about $+0.114$~e/Ti.

As described in the next paragraph in more detail, even the interfacial Ni loses electrons. We reach all these conclusions by comparing with the $3d$ electronic occupation in bulk NNO, in qualitative agreement with other recent theoretical calculations ~\cite{Geisler:prb}. It should also be noted that this charge transfer behavior was also observed in the unrelaxed slab structure, indicating that the charge transfer is caused by the polar discontinuity, namely it is electronic in origin, not due to the lattice relaxation. Due to the polar discontinuity, an intrinsic electronic reconstruction was also observed at the STO/LAO interface \cite{Ohtomo:nature04}. To avoid such polar catastrophe, the charge transfer occurred from surface to interface, leading to the observed conductivity~\cite{Nakagawa:nm} and superconductivity~\cite{Brinkman:nm}.

For the (CCO)$_n$/(STO)$_4$ superlattices, we also observed a hole-doping effect at the CuO$_2$ surface layer similar to the electrons transferred to the Ti interface for the case NNO/STO, see Fig.~\ref{Fig2}(e). The magnitude of charge transfer is slightly smaller for CCO than in the (NNO)$_n$/(STO)$_4$ system: here the surface Cu losses electrons with a net change $-0.072$~e/Cu while the interface Ti attracts electrons with a net gain $+0.085$~e/Ti.

More specifically, let us quantitatively discuss the charge transfer at the Ni or Cu sites of each layer corresponding to the $4$ unit-cell NiO$_2$ or CuO$_2$ layers, with respect to their bulk properties. For the case of NNO/STO,
all Ni sites (and corresponding layers) lose electrons. Because electrons are lost, we use a negative sign in front.
From surface to interface, the lost electrons at each Ni are $-0.101$~e/Ni, $-0.023$~e/Ni, $-0.005$~e/Ni,
and $-0.012$~e/Ni, respectively.  This tendency towards charge transfer is also qualitatively consistent with previous GGA+$U$ calculations~\cite{Geisler:prb}. Both the electronic and atomic structures will affect the charge transfer in oxide interfaces~\cite{Chen:jpcm,Zhong:prx}. In fact, we found that the Hubbard $U$ on Ni sites would increase the charge transfer from the Ni surface, as discussed in Appendix C. Thus, all Ni layers lose electrons and those electrons move to the STO portion of the superlattice, primarily to the STO interfacial layer.

With respect to CCO/STO, the modifications in the number of electrons at the Cu sites from surface to interface
are $-0.072$~e/Cu, $-0.018$~e/Cu, $+0.004$~e/Cu, and $+0.018$~e/Cu, respectively. The plus sign in the last two
means that in the case of CCO  the layers close to the interface gain electrons, the opposite behavior as for the case of NNO. In summary, {\it for NNO the surface-Ni/interfacial-Ni/interfacial-Ti display hole/hole/electron doping, while
for CCO the surface-Cu/interfacial-Cu/interfacial-Ti display hole/electron/electron doping}. This different behavior may be due to the different bending bond direction of the interfacial NiO$_2$ and CuO$_2$ layers. The common factor is that in both cases the surface layer loses electrons that eventually are transfered in part to the
Ti atoms. This behavior is similar to that reported in the LAO/STO system~\cite{Okamoto:nature,Bert:np,Dikin:Prl,Pavlenko:Prb}.

\subsection{B. Electronic reconstruction}

To study the movement of electrons due to the interface and generation of a 2D electron gas~\cite{n4context}, we calculated the ``electron localization function'' (ELF) for both NNO/STO and CCO/STO~\cite{Savin:Angewandte}, quantity widely used within {\it ab initio} methods to characterize the electronic charge distribution. As shown in Fig.~\ref{Fig1}(d), there are extra electrons emerging at the NNO-STO interface leading to 2DEG characteristics. In addition, the electronic distribution located at the intermediate  NiO$_2$ layers -- between interface and surface -- resembles the NNO/STO interface, namely the 2DEG extends deep into the NNO region. For CCO/STO, we also observed the presence of extra electrons at the CCO/STO interface, but the rest of the CCO layers are not much affected.

\begin{figure*}
\centering
\includegraphics[width=0.96\textwidth]{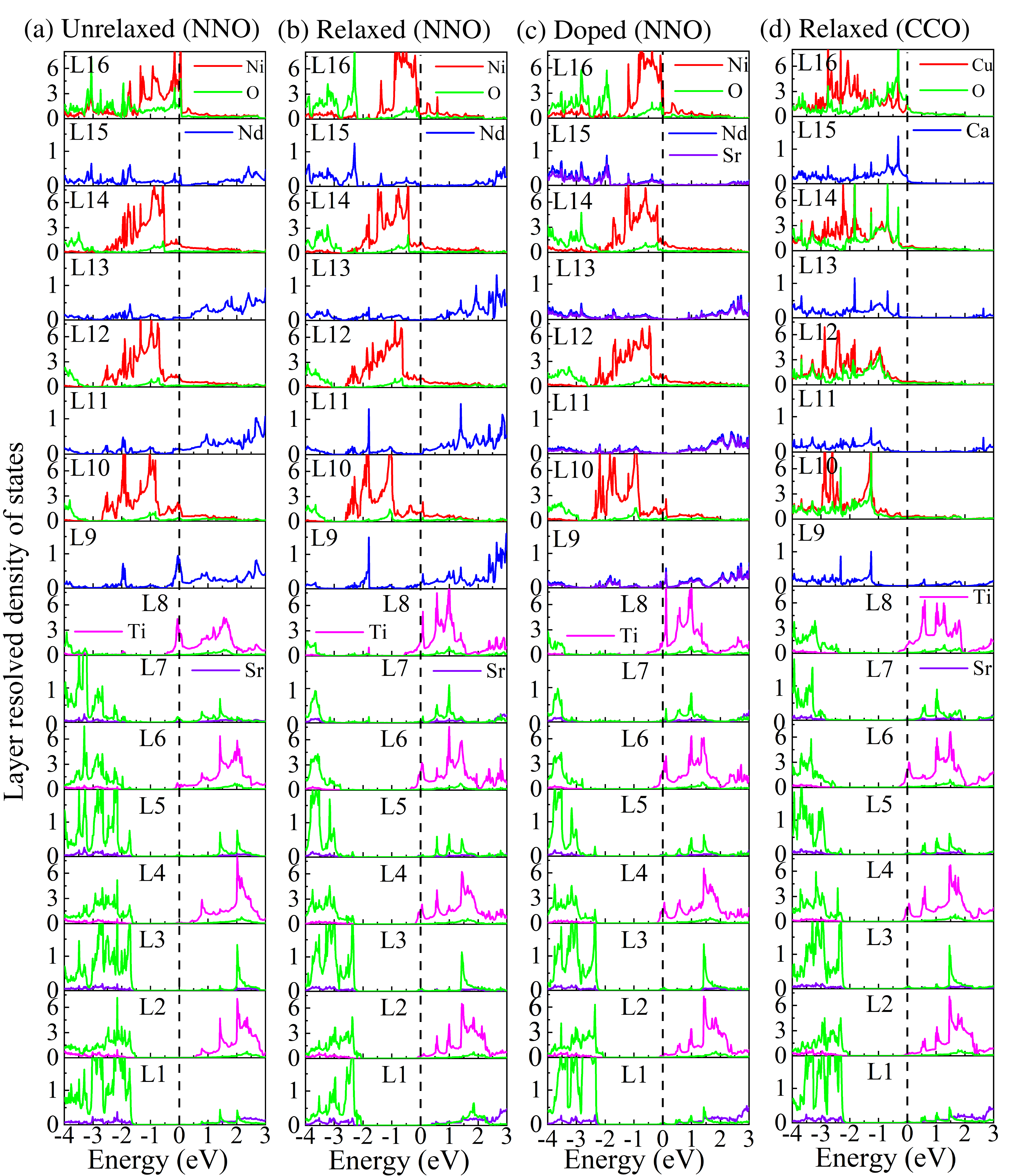}
\caption{(a-c) Electronic density-of-states (DOS) of each layer for the case (NNO)$_4$/(STO)$_4$, calculated via DFT. The Fermi level is shown with dashed vertical lines. Results presented are for (a) the unrelaxed slab model, (b) the relaxed atomic positions superlattice model, and (c) the $20\%$ hole-doped NNO model. For comparison, we also show (d) the DFT electronic DOS of each layer for the case (CCO)$_4$/(STO)$_4$ using the relaxed superstructure. The Fermi level is the dashed vertical line.}
\label{Fig3}
\end{figure*}

As shown in Fig.~\ref{Fig3}(a), for the case without lattice reconstruction, the active O-$2p$ orbitals shift closer to the Fermi level. This suggests that unoccupied oxygen $2p$ states may be involved in the electronic gas at the NiO$_2$ surface layer, supporting the pd hybridization picture (namely, oxygen is not so deep in energy that it decouples). The STO layers far from the interface display band insulator behavior consistent with bulk STO. However, at the two STO layers near NNO, the Ti $3d$ orbital becomes partially occupied, resulting in metallic STO layers. Our DOS results clearly show that a 2DEG emerges at the interface in the NNO films [see Fig.~\ref{Fig3}(a)].

By introducing the lattice relaxation, the electronic structures of both the interface and surface of NiO$_2$ are reconstructed and the $p-d$ hybridization strength is reduced due to the considerable O-Ni-O bending, in agreement with previous results~\cite{Geisler:prb,He:prb}, as displayed in Fig.~\ref{Fig3}(b). Furthermore, this panel indicates that the electrons transfer to the STO substrate. Overall, this suggests that the 2DEG at the interface is caused by the electronic reconstruction due to polar discontinuity. In LAO/STO systems, one
possible mechanism relies on the polar catastrophe that plays an important role in the formation of the two-dimensional electron gas, leading to superconductivity~\cite{Okamoto:nature,Nakagawa:nm,Nakagawa:nm}. This suggests that superconductivity in the NNO/STO of our focus may be related to as interfacial electron gas as well, as in LAO/STO.

To understand the effect of $20\%$ Sr-doping in the infinite-layer nickelate, we employed the virtual crystal approximation (VCA) widely used in the electronic structure
context~\cite{Bellaiche:Prb,Ramer:Prb}. As expected, hole doping can be understood using the rigid band picture as shown in Fig.~\ref{Fig3}(c), namely the existence of a 2DEG and hole doping at the surface are not qualitatively affected. Based on the analysis of the Ni $3d$ orbital population (not shown), the hole doping at the NiO$_2$ interface is primarily contributed by the Ni $d_{x^2-y^2}$ orbital.
Similarly as in Fig.~\ref{Fig3}(d), we also observed the electronic reconstruction and strong $p-d$ hybridization in the CCO layers. The hole-doping at the surface was contributed by the Cu $d_{x^2-y^2}$ orbital, as it occurs for the nickelate.

Furthermore, we also observed that the Ni $e_g$ orbital polarization in the infinite-layer nickelate films, defined as $P$ = [n($d_{ 3z^2-r^2}$)-n($d_{ x^2-y^2}$)]/[n($d_{ 3z^2-r^2}$)+n($d_{ x^2-y^2}$)], increases from interface Ni ($\sim$ $9\%$) to surface Ni ($\sim$ $27\%$), as in previous results~\cite{Geisler:prb}. We also calculated the Ni $e_g$ orbital polarization $P$ for the unrelaxed superstructure and here $P$ increases from interface Ni ($\sim$ $9.5\%$) to surface Ni ($\sim$ $25\%$), close to the values of the relaxed superstructure. Hence, this $e_g$ orbital polarization is caused by electronic effects, not structural. In addition, we observed that the Cu $e_g$ orbital polarization is smaller than the Ni $e_g$ orbital polarization: $P$ increases from interface Cu ($\sim$ $7\%$) to surface Cu ($\sim$ $15\%$).

\begin{figure}
\centering
\includegraphics[width=0.48\textwidth]{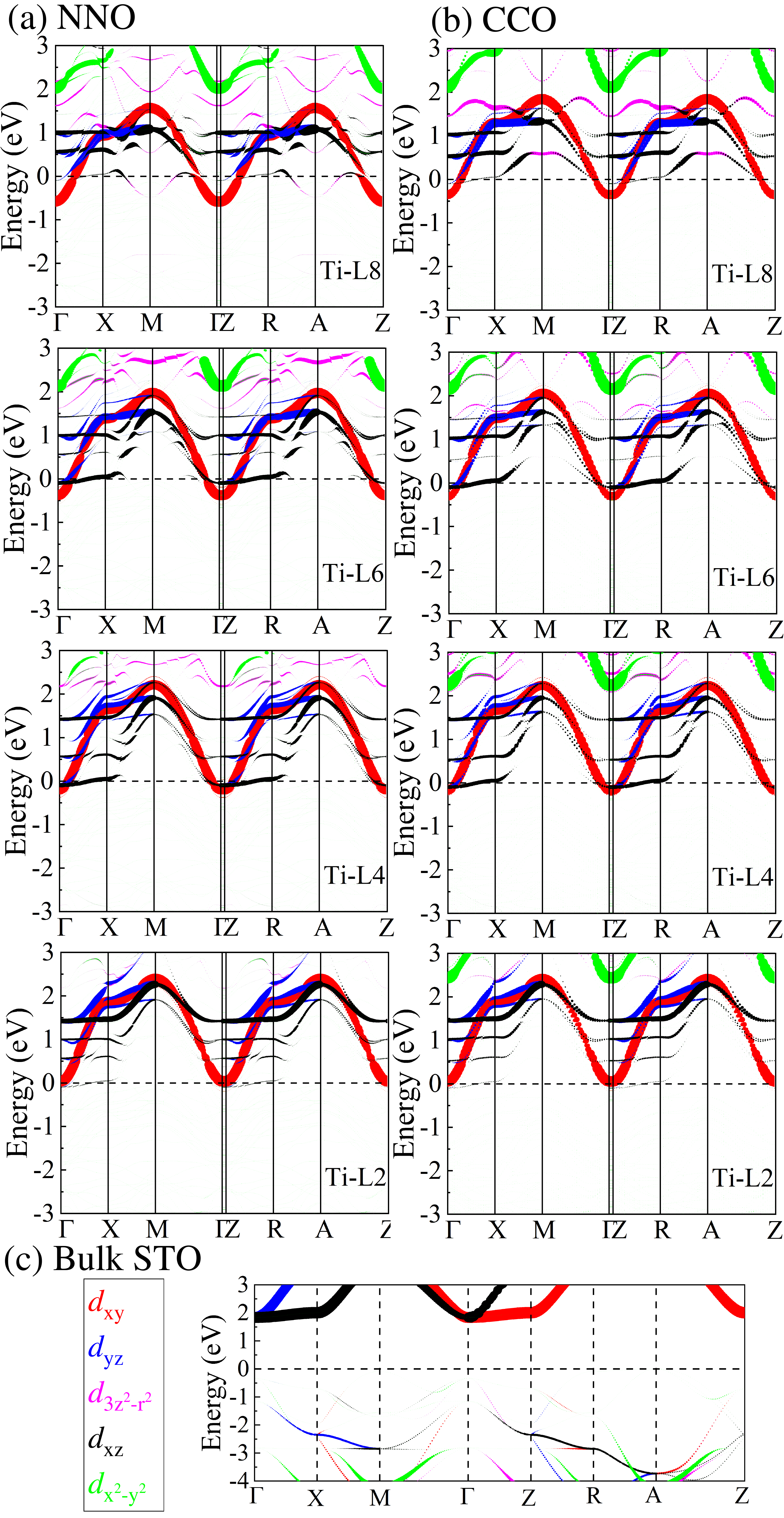}
\caption{Projected Ti band structure of the SrTiO$_3$ layers for the non-magnetic (NM) state. The Fermi level is shown with dashed horizontal lines. The weight of each Ti orbital is represented by the thickness of the lines. Note the $d_{xy}$ orbital,  particularly at $\Gamma$, is starting to be occupied because it is below the Fermi level. The electrons populating this portion of the Ti band arise from the Ni or Cu oxide films which, thus, become hole doped. Shown are (a) NNO/STO for various layers [label convention in Fig.~\ref{Fig1}~(a)] and (b) CCO/STO for various layers. For comparision, in (c) we show the Ti-bands of STO (bulk form) displaying the typical band-insulator gap and with an unoccupied $d_{xy}$ band, as discussed in the text.}
\label{Fig4}
\end{figure}

Next, let us discuss the charge transfer at the SrTiO$_3$ interface layers. As shown in Fig.~\ref{Fig4} [see panels (a,b) compared with bulk (c)\cite{bandgap}], the Ti $d_{xy}$ orbital starts to be occupied at the interface due to the charge reconstruction caused by the polar discontinuity. This is different from bulk STO where this orbital is unoccupied. More specifically, at the interfacial TiO$_2$ layer, Ti $3d_{xy}$ lowers its energy approximately by $-0.58$~eV and $-0.37$~eV at the $\Gamma$ point for NNO/STO and CCO/STO, respectively. This implies that electrons are
transfered to the STO substrate from the other component of the superlattice. Since STO is a wide band insulator, not a Mott insulator, the doping by electrons (holes) can be regarded as Fermi level shifts towards the conduction band (valence band). The electronic correlation on the Ti site would not change the charge transfer tendency but can increase the charge transfer values, as in the GGA+$U$ calculations in Ref.\cite{Geisler:prb}. In the previously widely studied case of the LAO/STO interface, also the STO Ti $3d_{xy}$ orbital plays an important role with regards to the interfacial ferromagnetism (FM) and superconductivity, induced by electronic reconstruction~\cite{Okamoto:nature,Bert:np,Dikin:Prl,Pavlenko:Prb}. As a consequence, our results indicate that Ti $3d_{xy}$ becomes active at the interface and suggest that this
orbital may be important for superconductivity in NNO as well, similarly as in LAO/STO. Experimental work is needed to confirm our conjecture but our effort points toward many similarities between LAO/STO and NNO/STO-CCO/STO, as well as a non-passive role of the STO component. In CCO/STO, a theory indicates that the introduction of extra oxygen ions at the CCO/STO interface is important for superconductivity~\cite{Castro:prl}.

\subsection{C. Magnetic reconstruction}

Theoretical efforts strongly suggest that N\'eel (G-type) antiferromagnetism (AFM) has the lowest energy on $R$NiO$_2$ ($R$= La/Nd) bulk \cite{Lee:prb,Ryee:prb,Zhang:prr,Leonov:arxiv}. However, magnetization and neutron powder diffraction revealed no long-range magnetic ordering in these compounds, although short-range spin correlations are possible \cite{Hayward:jacs,Hayward:ssc}. Other explanations have been proposed~\cite{Li:cm}.

To better understand correlation effects and the possible magnetism at the Ni ions, here we use an effective
Hubbard coupling ($U_{\rm eff}$ = 4 eV)~\cite{Dudarev:prb,Geisler:prb} assuming G-AFM magnetic order in the NiO$_2$ layers of the superlattice. It should be noted that the physical description will not change by considering magnetism and electronic correlations on the Ni sites, i.e. the ionic reconstruction, charge transfer, and the strong occupation of the 3$d_{\rm xy}$ state are robust conclusions of our effort.

\begin{figure}
\centering
\includegraphics[width=0.48\textwidth]{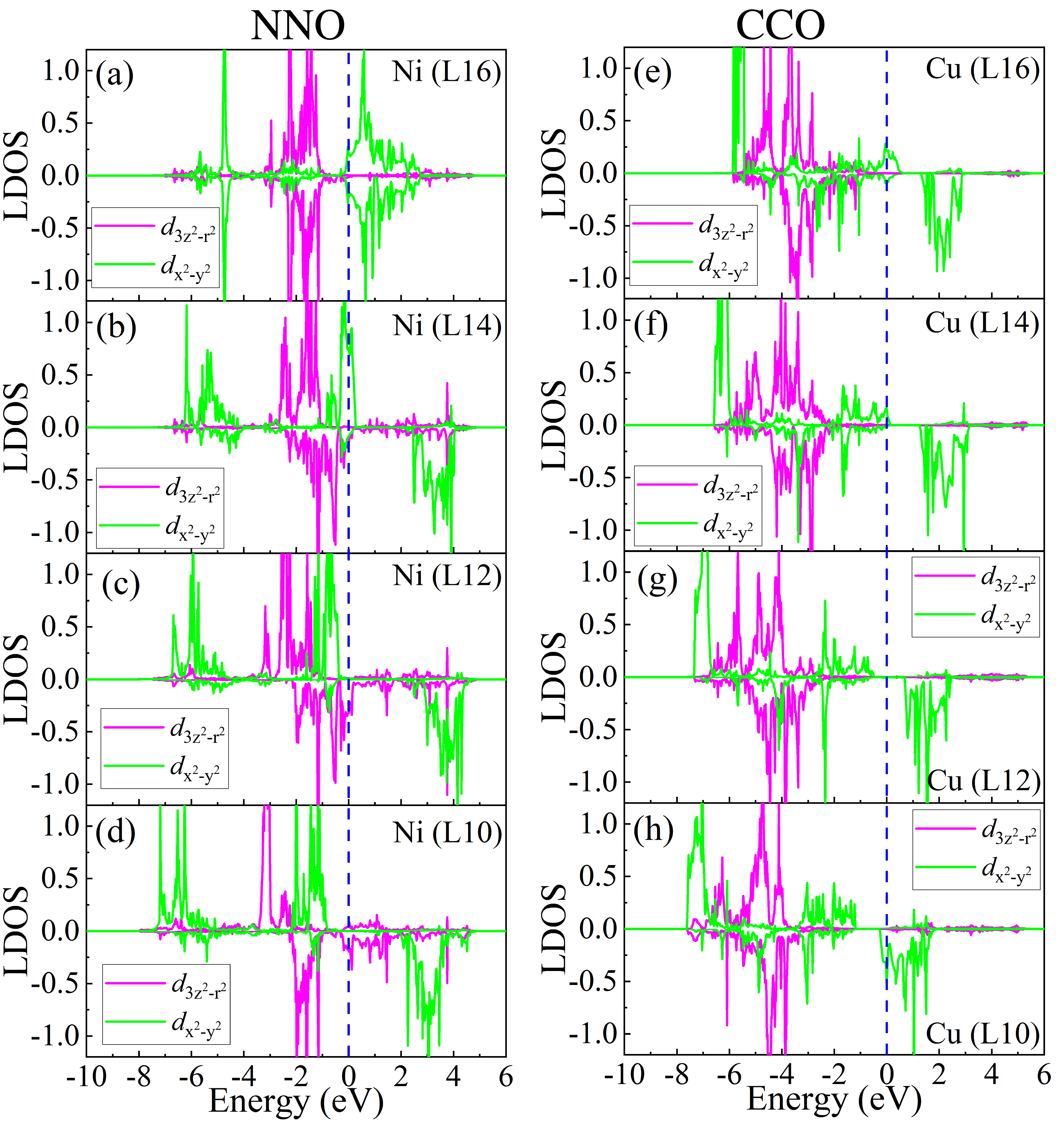}
\caption{ Ni and Cu projected local DOS corresponding to the $e_{\rm 2g}$ orbital of (NNO)$_4$/(STO)$_4$ and (CCO)$_4$/(STO)$_4$ assuming a G-AFM type magnetic configuration. The Fermi level is indicated with dashed lines. Note that we only show the local DOS of one Ni or Cu site at each layer.}
\label{Fig5}
\end{figure}

Interestingly, we observed a strong magnetic reconstruction at the NdNiO$_2$ surface layer, see Fig.~\ref{Fig5}(a), as compared to the internal layers, such as Fig.~\ref{Fig5}(d). The surface is basically non-magnetic. Furthermore, based on our DFT results with G-AFM magnetism and electronic correlation, we estimate that the surface Ni lost approximately $-0.129$~e/Ni while the interface Ti gained about $0.129$~e/Ti, as compared with the $3d$ electronic occupation in bulk. Thus, the tendency for charge transfer discussed in Fig.~\ref{Fig2}(b) was found not to change by introducing the magnetism and electronic correlation.

By studying the DOS in Figs.~\ref{Fig5} (b-d), we found that the magnetic moments of Ni are primarily contributed by the $d_{x^2-y^2}$ orbital since the electronic occupation of spin-up and spin-down of $d_{3z^2-r^2}$ are almost equal. In the two NNO layers near STO, $d_{x^2-y^2}$ displays a Mott gap
while the $d_{3z^2-r^2}$ orbital remains metallic, which is close to the theoretical results obtained for bulk NNO~\cite{Lechermann:prb}. As shown in Fig.~\ref{Fig5}(b), comparing with the L10 and L12 layers, the $3d_{x^2-y^2}$ orbital began to shift across the Fermi level, indicating that the Ni $3d_{x^2-y^2}$ orbital of the L14 layer would also lose some electrons. The hole-doping at the surface reduces the electronic occupation of Ni $3d_{x^2-y^2}$, leading the original Ni$^{\rm 1+}$ ($d^9$) to change towards Ni$^{\rm 2+}$ ($d^8$), although there is a partial electronic occupation of $3d_{x^2-y^2}$ [see Figs.~\ref{Fig5}(a) and (d)].

Furthermore, we also considered the case of $U_{\rm eff}$ = 6 eV ~\cite{Xiang:prb} with the G-AFM magnetic configuration in each CuO$_2$ layer for the CCO/STO structure. In Fig.~\ref{Fig5}(e), we calculated that the surface has a moment $0.55$ $\mu_{\rm B}$/Cu. The central CuO$_2$ layers, between surface and interface, still remain insulating while the surface and interfacial CuO$_2$ layers become metallic due to hole and electron doping, respectively. Different from NNO/STO, the Cu $3d_{3z^2-r^2}$ orbital displays doubly occupied behavior and for this reason this orbital is not active for the Cu layers, as widely accepted. In this case, the different behavior of $d_{3z^2-r^2}$ between nickelate and cuprate films establish another qualitative difference between the two cases.

\section{IV. Conclusions}

In summary, we systematically studied nickelate and cuprate films grown on an STO substrate by using first-principles calculations. We observed a strong lattice reconstruction in both NNO/STO and CCO/STO systems. In the case of NNO/STO, the surface and interfacial NiO$_2$ layers present opposite displacements, although the
net effect still lead to a globally NNO polar film due to different displacement values. For the case of CCO/STO, the surface and interfacial CuO$_2$ layers were distorted in the same direction, again leading to a globally CCO polar film. Moreover, the atomic reconstruction was basically independent of the thickness of the NNO or CCO films, at least up to the 8 layers studied here. We also observed a strong electronic reconstruction for the two systems caused by the growing internal electric field (polar catastrophe), as in other superlattices before. In the NNO/STO system, a 2DEG formed at the interfacial layer between NNO and STO, extending over several layers. For the CCO/STO system, on the other hand, the 2DEG was more sharply localized at the interface between CCO and STO. Furthermore, a strong occupation of the Ti $3d_{xy}$ states was observed at the TiO$_2$ interface for both NNO/STO and CCO/STO. In addition, the surface $e_g$ orbital polarization $P$ of Ni was found to be stronger than the surface $e_g$ orbital polarization $P$ of Cu. By introducing magnetism and electronic correlation, the $d_{3z^2-r^2}$ orbital of Ni, which is near insulating in the absence of Hubbard $U$, remarkably becomes itinerant while the $d_{3z^2-r^2}$ orbital of Cu remains doubly occupied. For this reason, we believe that the NNO grown on STO requires a two-orbital description as often employed in nickelates,  while CCO admits the canonical one-orbital formulation standard for cuprates. In addition, we also observed a strong magnetic reconstruction at the NdNiO$_2$ surface to vacuum layer where magnetism is basically suppressed.

\section{Acknowledgments}
E.D., Y.Z., L.F.L., W.H., and A.M. are supported by the U.S. Department of Energy (DOE), Office of Science, Basic Energy Sciences (BES), Materials Sciences and Engineering Division. S.D. is supported by the National Natural Science Foundation of China (Grant Nos. 11834002 and 11674055). All the calculations were carried out at the Advanced Computing Facility (ACF) of the University of Tennessee Knoxville.
\begin{appendices}

\section{APPENDIX}

\subsection{A. Electronic structure of bulk NdNiO$_2$ and CaCuO$_2$}

\begin{figure}
\centering
\includegraphics[width=0.48\textwidth]{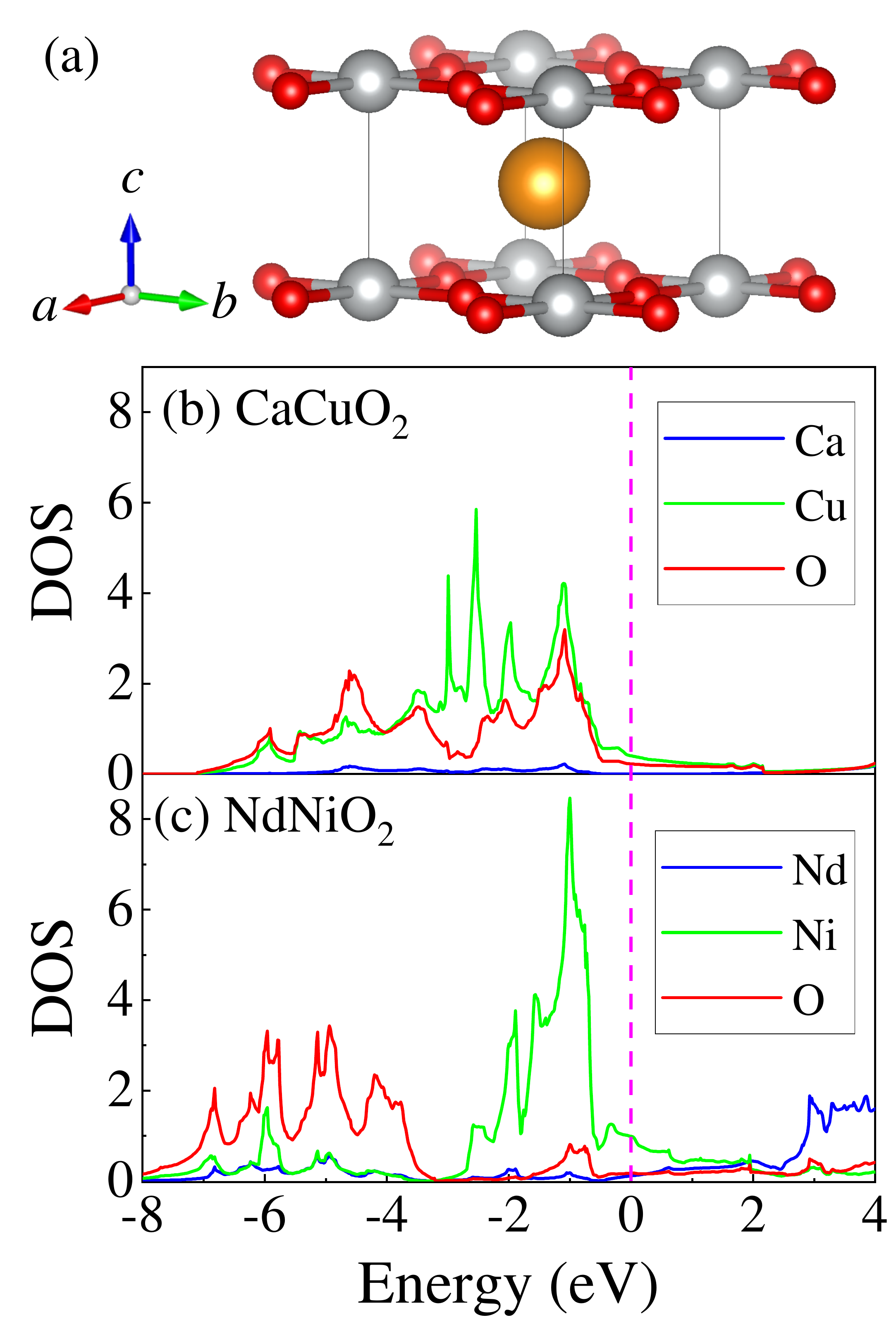}
\caption{(a) Schematic crystal structure of $ABO_2$ (electronic density $n = 9$) with the convention: Orange = Sr or Nd; Gray = Cu or Ni; Red = O. (b-c) DOS near the Fermi level using the non-magnetic states for (b) CaCuO$_2$ and (c) NdNiO$_2$, respectively.}
\label{Fig6}
\end{figure}

\begin{figure}
\centering
\includegraphics[width=0.48\textwidth]{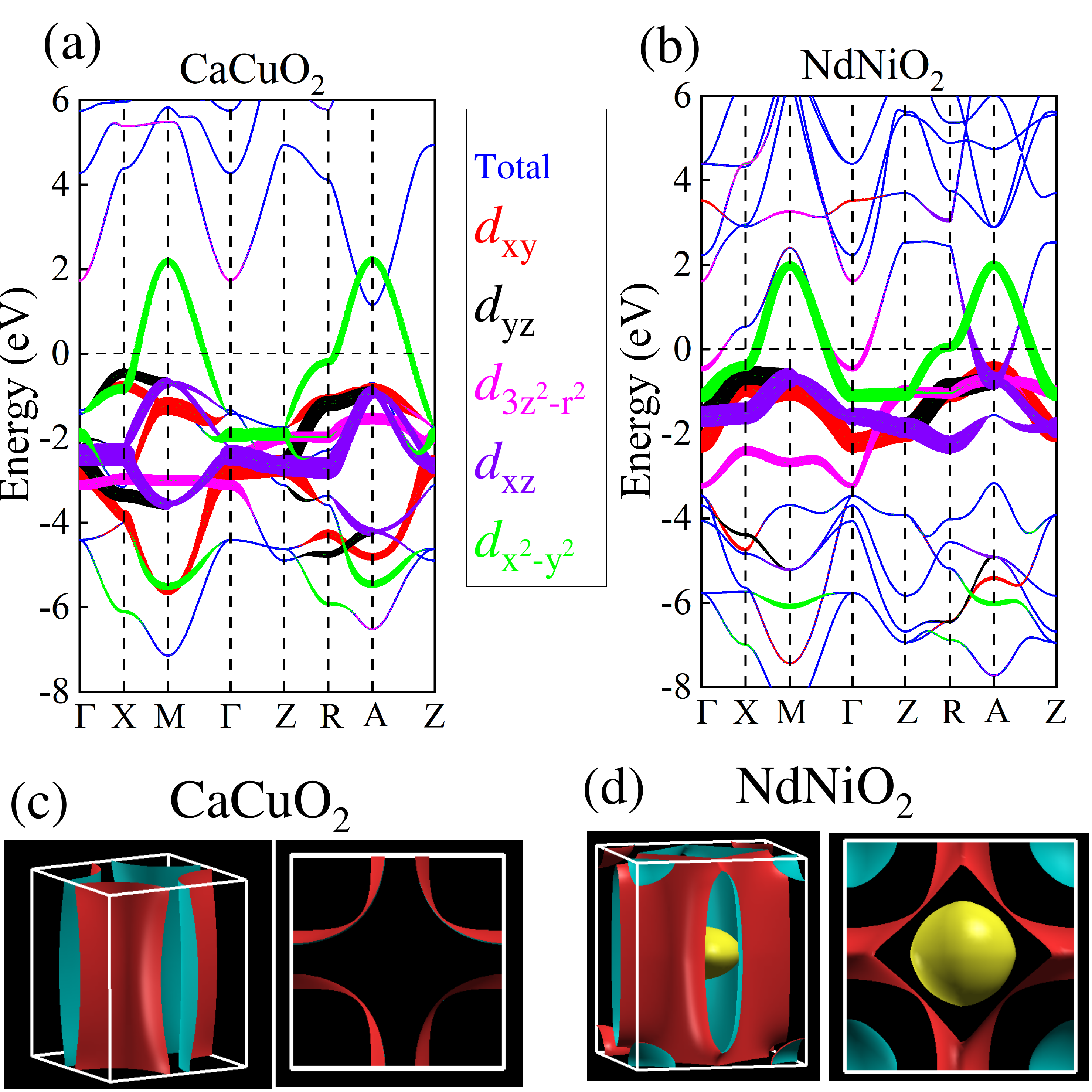}
\caption{ (a-b) Projected band structures of CaCuO$_2$ and NdNiO$_2$ for the non-magnetic state, respectively. The Fermi level is shown with horizontal
dashed lines. The weight of each orbital is represented by the thickness of the lines. (c-d) Fermi surface of CaCuO$_2$ and NdNiO$_2$, respectively.}
\label{Fig7}
\end{figure}

NdNiO$_2$ forms a $P4/nmm$ tetragonal crystal structure, similar to CaCuO$_2$ [see Fig.~\ref{Fig6}(a)]. As shown in Fig.~\ref{Fig6}(b) and (c), the electronic density near the Fermi level is mainly contributed by the Cu or Ni atoms. The main difference is that there is a strong pd hybridization in CaCuO$_2$ but this hybridization is quite weak in NdNiO$_2$ (the latter is contrary to the results for NNO/STO shown in the main text where the hybridization $p-d$ is larger). In the case of Cu, only one Cu $d_{x^2-y^2}$ band crosses the Fermi energy. However,  there two bands crossing the Fermi level for  nickelates: Ni $d_{x^2-y^2}$ and Nd $d_{3z^2-r^2}$, the latter hybridized with $d_{yz}$, as displayed in Fig.~\ref{Fig7}(a-b). The $3d_{x^2-y^2}$ orbital is more dispersive along the $xy$ plane than along the $z$ direction due to the Cu-O or Ni-O antibonding character. The O $2p$ bands of NdNiO$_2$ extend over a broad range of energy from $-10$ eV to $2$ eV indicating that the charge transfer energy is quite larger than for CaCuO$_2$. Another interesting result is that the $3d_{3z^2-r^2}$ orbital of Ni is more broad than the $3d_{\rm 3z^2-r^2}$ orbital of Cu. This implies that the $3d_{\rm 3z^2-r^2}$ orbital of Ni may display different behavior as the same orbital in Cu. Based on the Wannier fitting using the WANNIER90 packages ~\cite{Mostofi:cpc}, we estimate the charge-transfer energy $\Delta$ = $\varepsilon_d$ - $\varepsilon_O$  to be  $1.7$~eV for CaCuO$_2$ and $4.2$~eV for NdNiO$_2$.

\subsection{B. Compressive strain on the epitaxial constraint to STO}

\begin{figure}
\centering
\includegraphics[width=0.48\textwidth]{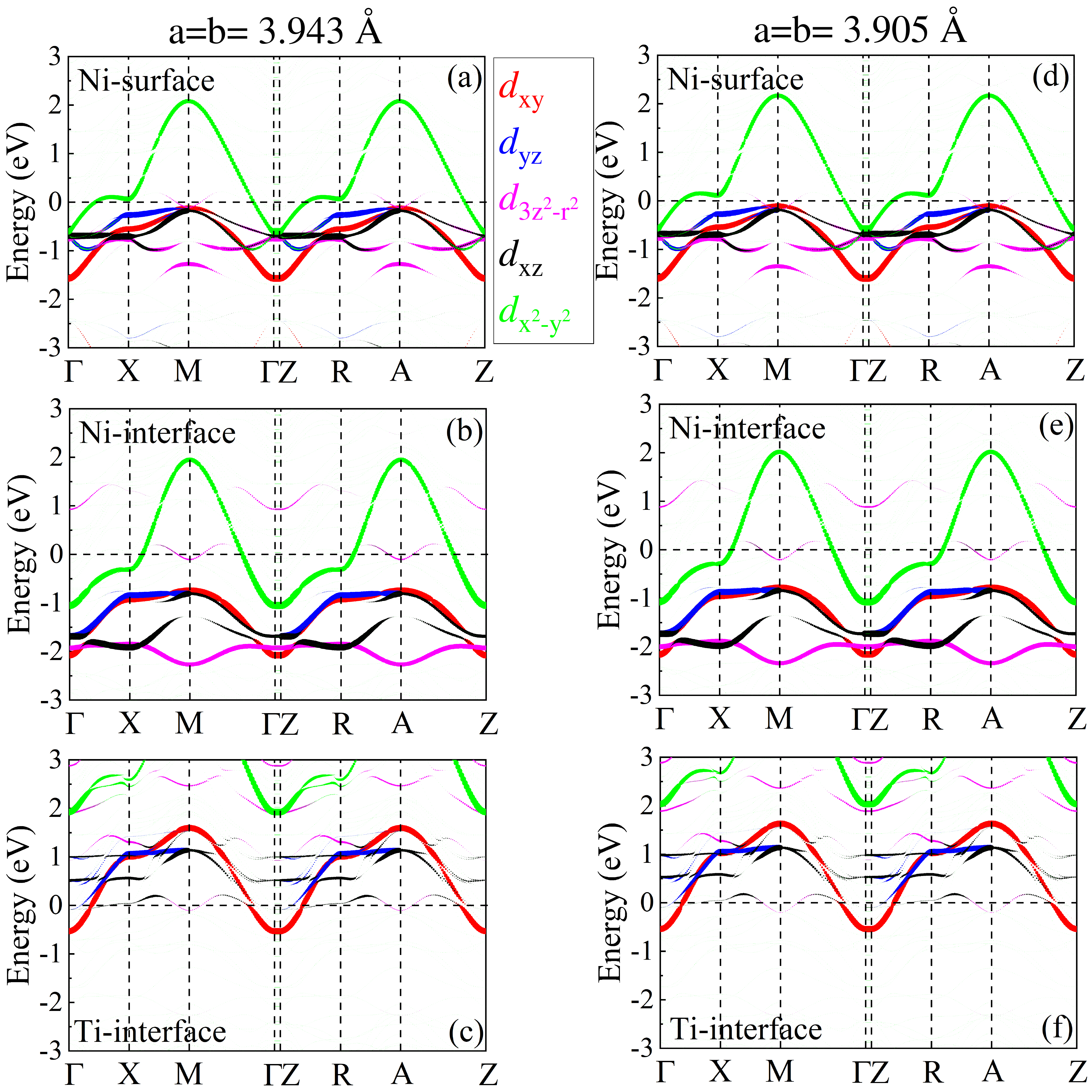}
\caption{Projected band structures of Ni and interface Ti for the case (NNO)$_2$/(STO)$_4$ using $a=b=3.905$~\AA ~and $a=b=3.943$~\AA, respectively. The Fermi level is shown with horizontal dashed lines. The weight of each orbital is represented by the thickness of the lines.}
\label{Fig8}
\end{figure}

Our relaxed crystal lattice constants of cubic STO are $a=b=c=3.943$~\AA. If we fix two of them as $a=b=3.943$~\AA ~without polar distortion, then the relaxed lattice constant of the $c$-axis becomes $4.032$~\AA ~and the strained STO is still paraelectric. However, if we introduce a polar distortion along the $c$-axis in a fixed structure, the energy becomes lower than the strained STO without polar distortion. In this case, the system then changes to a ferroelectric tetragonal phase. Since the optimized in-plane lattices of STO ($a=b=3.943$~\AA) are slightly larger than the experimental values ($3.905$~\AA), it would generate a moderate compress strain. To better understand this point, we also calculated the two layer NdNiO$_2$ slab model by fixing the in-plane lattice to the optimized STO ($a=b=3.943$~\AA). After a full structural optimization relaxation, the bending behavior of the NiO$_2$ layers do not change. The corresponding Ni ions still move towards {\it different} directions with displacements $\Delta= 0.279$ and $-0.169$~\AA, quite close to the results of $n = 2$ NNO/STO with fixed experimental in-plane lattice constants ($\Delta= 0.291$ and $-0.184$~\AA). Furthermore, the interfacial Ti ions of the STO substrate still show ferroelectric-like displacements away from the NNO/STO interface with values $\Delta= -0.291$ and $-0.145$~\AA, respectively, which is in closer agreement with the distortions of Ti for a fixed $a=b=3.943$~\AA ~($\Delta= -0.301$ and $-0.157$~\AA). In addition, the hole-doping at the interface, charge transfer behavior, existence of a 2DEG, electronic occupation of the Ti $3d_{xy}$ orbital, and other physical properties are consistent with the NNO/STO ($n = 2$) slab using fixed experimental STO lattice constants. For the benefit of the readers, we compared the band structures of Ni, at surface and interface, and interface Ti for the two cases, as shown in Fig.~\ref{Fig8}.

\end{appendices}

\subsection{C. Charge transfer with $U_{\rm eff}$}

\begin{figure}
\centering
\includegraphics[width=0.48\textwidth]{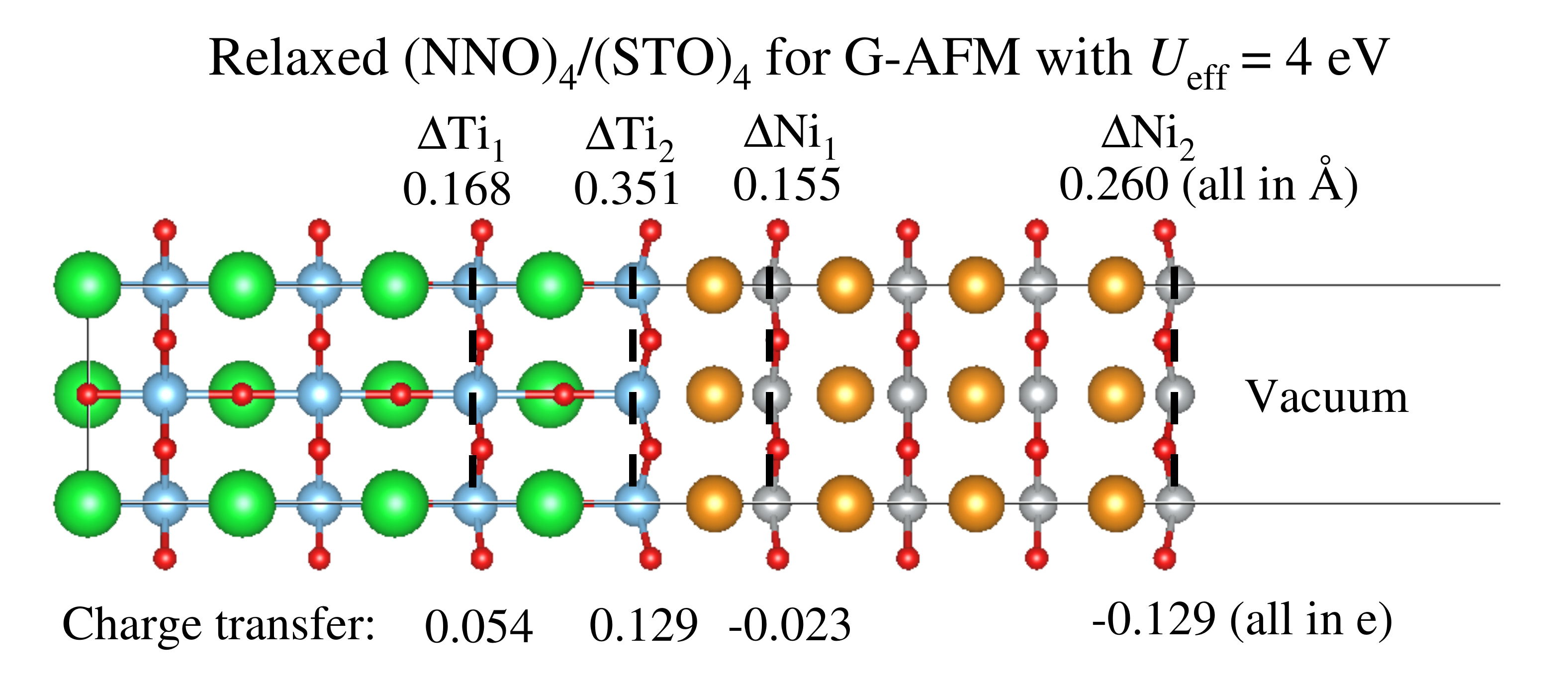}
\caption{Schematic relaxed slab model of (NNO)$_n$/(STO)$_4$ for G-AFM order ($U_{\rm eff}$ = 4 eV) with the convention: Orange = Nd or Ca; Gray = Ni or Cu; Red = O; Green = Sr; Blue = Ti. Here, we fixed $a=b=3.905$~\AA. We considered two undistorted STO layers (namely with atomic positions fixed) to simulate the effects of the rest of the STO substrate.}
\label{Fig9}
\end{figure}

To better understand the charge transfer, we performed DFT+$U_{\rm eff}$ calculations using a (NNO)$_4$/(STO)$_4$ slab. On one hand, we fully relaxed the structure with the G-AFM configuration by using $U_{\rm eff}$ = 4 eV on Ni sites, since both electronic and atomic structure will affect the charge transfer. As shown in Fig.~\ref{Fig9}, both the surface and interfacial Ni ions move in {\it different} directions, with displacements $\Delta= 0.260$ and $-0.155$~\AA, which is slightly reduced by comparing with results of DFT for the NM state ($\Delta= 0.296$ and $-0.189$~\AA). Based on our calculations, from surface to interface, the lost electrons at each Ni are $-0.129$ e/Ni, $-0.031$ e/Ni, $-0.001$ e/Ni, and $-0.023$ e/Ni, respectively, while the Ti$_1$ and Ti$_2$ captured electrons are
about $0.054$ e/Ti and $0.129 $ e/Ti. In contrast, the results of DFT of NM are -0.101 e/Ni, -0.023 e/Ni, -0.005 e/Ni, and -0.012 e/Ni, respectively.

On the other hand, we also added $U_{\rm eff} = 4 eV$ directly to the optimized structure from pure GGA (namely, keeping the atomic distortions unchanged) since the atomic structure only slightly changed in the DFT+$U_{\rm eff}$ calculations. In this case, we estimate that the surface Ni lost approximately $-0.141$~e/Ni while the interface Ti gained about $0.126$~e/Ti, as shown in Table~\ref{Table1}. Hence, the charge transfer is enhanced by introducing $U_{\rm eff}$.

\begin{table}
\centering\caption{The charge transfer of Ni and Ti atoms for different $U_{\rm eff} = 4$ eV based on the optimized GGA structure, compared with the corresponding bulk values. The subindex "int" ("sur") denotes interface (surface).}
\begin{tabular*}{0.48\textwidth}{@{\extracolsep{\fill}}llllc}
\hline
\hline
  & $Ti_1$ &  $Ti_2$  & $Ni_{\rm sur}$ & $Ni_{\rm int}$\\
\hline
Without $U_{\rm eff}$  & 0.065  &0.114 & -0.012 & -0.101  \\
$U_{\rm eff}$ = 4 eV   & 0.06   &0.114 & -0.009 & -0.141  \\
\hline
\hline
\end{tabular*}
\label{Table1}
\end{table}


\begin{references}
\bibitem{bandgap}The band gap of STO is $1.86$~eV based on GGA calculations. This value is smaller than the experimental value ($\sim$ 3.2 eV) \cite{Brinkman:nm}, since the DFT technique usually underestimates the band gap of ABO$_3$ perovskite materials. It should be noted that hybrid-exchange correlation functionals, such as B3LYP, allows to achieve an excellent agreement with experiments for the band gaps of related ABO$_3$ perovskites, their (001) surfaces, and heterostructure~\cite{Eglitis:jsc}. The hybrid-exchange correlation functional provides only a correction for the band gap, and does not change other physical properties. However, the scope of this publication is to focus on the physical properties at the interface. Hence, the GGA calculations are good enough to describe this superlattice structure. While, as expressed above, it is well known that the DFT technique usually underestimates the band gap of ABO$_3$ perovskite materials, hybrid-exchange correlation functionals, such as B3LYP, allow to achieve an excellent agreement with experiments for the band gaps of related ABO$_3$ perovskites, their (001) surfaces, and heterostructures \cite{Eglitis:jsc}.
\bibitem{chargetransfer} Here, the charge transfer was calculate by comparing the difference of occupied number of $3d$ orbitals between superlattice and corresponding bulk structure.
\bibitem{n4context} Since we verified that the film thickness of NNO or CCO does not change the main physical conclusions of our work, we primarily discussed our main results using the (NNO)$_4$/(STO)$_4$ or (CCO)$_4$/(STO)$_4$ heterostructures, corresponding to a film thickness of $4$ unit cells.
\end{references}
\end{document}